\pdfoutput=1
\documentclass[lettersize,journal]{IEEEtran}
\usepackage{graphicx}
\usepackage{cite}
\usepackage{picinpar}
\usepackage{amsmath}
\usepackage{url}
\usepackage{flushend}
\usepackage[latin1]{inputenc}
\usepackage{colortbl}
\usepackage{soul}
\usepackage{multirow}
\usepackage{pifont}
\usepackage{color}
\usepackage{alltt}
\usepackage[hidelinks]{hyperref}
\usepackage{enumerate}
\usepackage{siunitx}
\usepackage{breakurl}
\usepackage{epstopdf}
\usepackage{pbox}
\usepackage{subcaption}
\usepackage{tabularx}

\usepackage{multirow}
\usepackage{float}
\usepackage{stfloats}
\usepackage{amsmath,amsfonts,amssymb,amscd}
\usepackage{hyperref}
\usepackage{mathptmx}
\usepackage{textgreek}
\usepackage{bm}
\usepackage{caption}
\usepackage{subcaption}
\usepackage{flushend}
\usepackage{cuted}
\usepackage{titlesec}
\titlespacing{\subsection}{0pt}{8pt}{5pt} 
\titlespacing{\subsection}{0pt}{5pt}{3pt} 

\begin{document}
\title{Y-Configuration Active Bridge (YAB) Converter: A DAB-Type Single-Stage Isolated Three-Phase AC-DC Converter with Simple Sinusoidal Control}

\author{
	\vskip 1em
	\vspace{-10pt}
	    Mafu Zhang, 
	    Huanghaohe Zou,
        Saleh Farzamkia,
        Zibo Chen,
        Chen Chen
        and Alex Q. Huang, \emph{Fellow, IEEE}
        \vspace{-25pt}
        
        \thanks{The authors are with the Semiconductor Power Electronics Center, The University of Texas at Austin, Austin, TX 78712 USA}
}

\maketitle

\begin{abstract}
This paper reviews commonly used three-phase isolated AC-DC converters and introduces a novel Y-configuration Active Bridge (YAB) converter for single-stage isolated AC-DC power conversion. The proposed YAB addresses the limitations of multi-stage designs by eliminating bulky electrolytic capacitor banks and input boost inductors, thereby enabling a simplified start-up process without the risk of inrush current. It retains the advantages of single-stage AC-DC Dual Active Bridge (AC-DC DAB) converters while significantly reducing control complexity. Across its entire operating range, the YAB achieves low total harmonic distortion, maintains relatively low current stress, and exhibits excellent soft-switching performance. The operating principle of the proposed converter is detailed, and an equivalent circuit model is presented. System performance is evaluated using a numerical Fast Fourier Transform (FFT) model. To validate the performance of the proposed converter, a three-phase 6 kW, 480 V YAB prototype is designed and tested in the laboratory. Experimental results demonstrate a maximum efficiency of 97.1\%.

\end{abstract}

{}

\definecolor{limegreen}{rgb}{0.2, 0.8, 0.2}
\definecolor{forestgreen}{rgb}{0.13, 0.55, 0.13}
\definecolor{greenhtml}{rgb}{0.0, 0.5, 0.0}

\section{Introduction}\label{sect:intro}
\IEEEPARstart{T}{hree-phase} isolated AC-DC converters are essential in high-power applications, such as hybrid AC-DC microgrids, AC-DC solid-state transformers (SSTs), battery energy storage systems, and EV chargers. Galvanic isolation is crucial in these systems to mitigate ground loops, enable voltage scaling, enhance fault tolerance, and support specific grounding configurations \cite{7151820}. Various isolated AC-DC converter topologies have been proposed in the literature, most of which can be broadly categorized into single-stage and two-stage power conversion architectures, as illustrated in Fig.\ref{fig:comparison}.

\subsection{Two-Stage Methods}
The conventional approach for three-phase AC-DC power factor correction (PFC) and galvanic isolation uses a two-stage topology. In this method, the PFC stage first rectifies the grid voltage to produce a relatively stable DC-link voltage, which is subsequently processed by the isolated DC-DC stage to deliver the desired load voltage.

\begin{figure*}[!t]
    \centering
    \vspace{-10pt} 
    \includegraphics[width=0.85\linewidth]{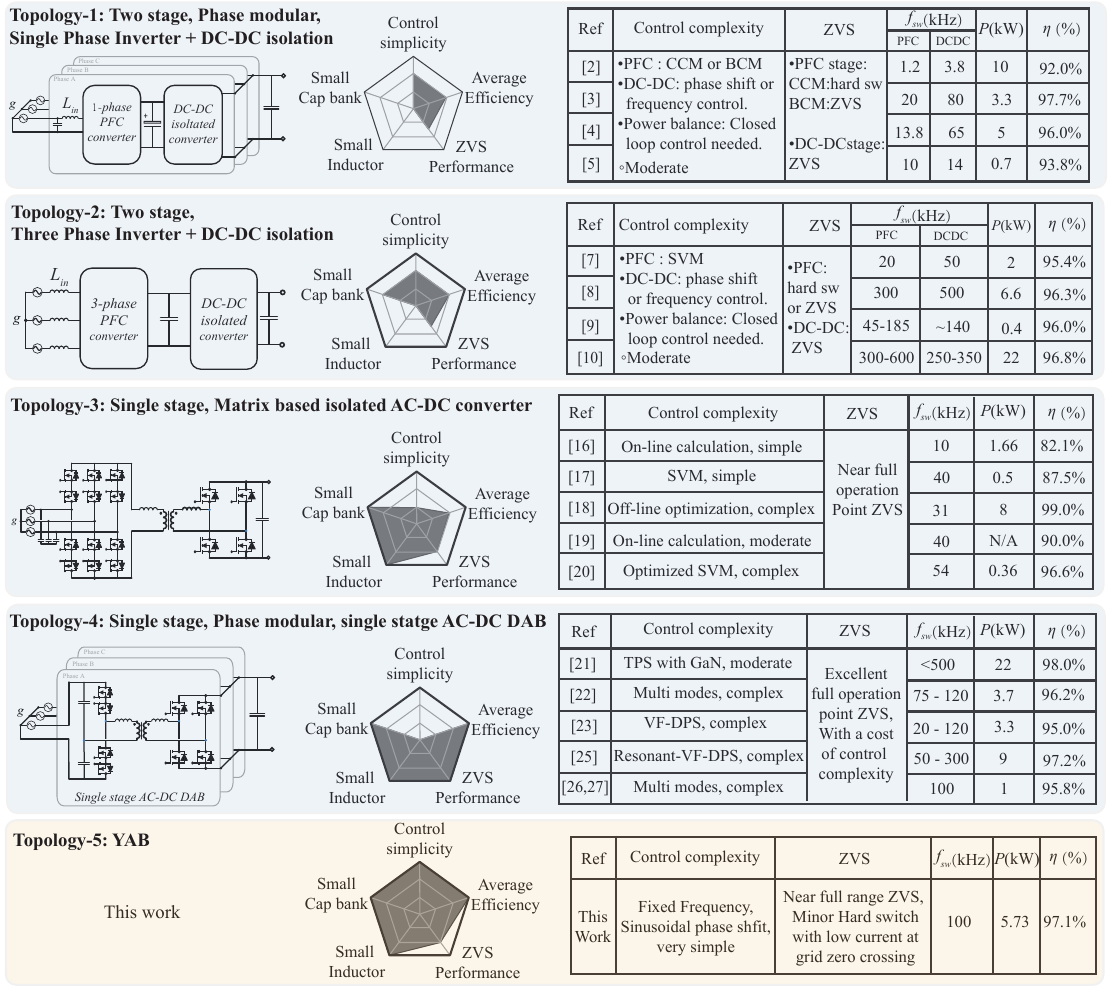}
    \caption{Comparison of existing three-phase isolated AC-DC PFC topologies}
    \vspace{-10pt} 
    \label{fig:comparison}
\end{figure*}

A phase-modular example of this approach is $Topology\text{-}1$, shown in Fig.\ref{fig:comparison} \cite{6678214,7762915,6269110,6975154}. This topology is commonly employed in AC-DC SSTs due to its phase-modular design, which simplifies phase-to-phase insulation by enabling flexible spacing to meet clearance and creepage requirements \cite{6678214}. The main drawback of this topology is the need for bulky electrolytic capacitors to suppress the double-frequency ripple in the intermediate DC-link voltage. These capacitors not only reduce system power density but also limit service life due to their finite operational lifespan. Additionally, they introduce inrush currents during startup, necessitating the inclusion of soft-start circuits with relays \cite{10131324}.

In applications where interphase insulation is less critical, monolithic conversion ($Topology\text{-}2$) \cite{7973139,8377992,6687217,GibumYu24} is favored. Compared to $Topology\text{-}1$, this topology reduces the intermediate DC-link capacitance due to the constant current output of the PFC stage. Although the required capacitance is smaller, it must still be sufficiently large to prevent voltage overshoot during transients caused by power flow imbalances \cite{8025412,7362206}. Consequently, electrolytic capacitors are still widely used, perpetuating the same limitations as $Topology\text{-}1$. Moreover, both topologies require substantial grid-side filter inductance, which is often bulky in high-power applications \cite{7762915,1628997}.

Typically, the hard-switching PFC and the ZVS DC-DC stages in these topologies achieve peak efficiencies exceeding 98.5\%. However, since the peak efficiencies of the two stages do not occur at the same operating point, the overall system efficiency is generally below 96\%. While advanced ZVS control techniques \cite{8377992,8502079} and GaN-based devices \cite{9319851} can further enhance efficiency and switching frequency, they often come at the expense of increased control complexity and higher costs.

\subsection{Single-Stage methods}

Single-stage power conversion is a promising solution for achieving both high efficiency and high power density. It eliminates the need for large electrolytic capacitors and filter inductors. In addition to these benefits, single-stage systems are more cost-effective, as each additional stage increases the total semiconductor die area required for a given power design.

Among single-stage structures, matrix-based isolated AC-DC converter and phase-modular single-stage AC-DC DAB converter - $Topology\text{-}3$ and $Topology\text{-}4$ in Fig. 7 - have recently received significant attention \cite{8573905,7585108,8704941,8888240,10143345, 8094272,6671445,7377097,9583867,9796024,9356460,9351779}. Beyond the aforementioned benefits, their DAB-type characteristics inherently enable stability during grid voltage ride-through events, even with open-loop or slow closed-loop control \cite{9807024,8507672}. This makes them particularly well-suited for grid-following applications such as load-shifting battery energy storage systems (BESS), Electric vehicle (EV) chargers, and AC-DC SSTs.

$Topology\text{-}3$ consists of an AC-side direct matrix converter, a high-frequency (HF) transformer, and a DC-side full bridge. It was originally proposed in \cite{5589710} as a V2G interface with bidirectional power flow. Subsequent studies \cite{8888240,10143345} enhanced its efficiency by employing optimized modulation methods. Furthermore, \cite{8704941} demonstrated a remarkable peak efficiency of 99\% through ZVS/ZCS modulation, which relies on offline optimization to minimize conduction losses.

The $Topology\text{-}4$ converter features a phase-modular design comprising three single-phase single-stage AC-DC DAB converters. First introduced in \cite{299007}, it employs either bidirectional switch bridges or active unfolding bridges to rectify the grid voltage into a half-sine waveform, then performs the unity power factor conversions with the DC-DC DAB \cite{6165948}. This topology achieves full-operation-range ZVS and high efficiency by carefully calculating multiple phase shifts and switching frequency. Typical modulation methods include optimization-based multi-mode transition \cite{6671445,9356460} or single-mode operations with wide-range varying switching frequency \cite{7377097,9583867}. Further enhancements, such as resonant operation \cite{9796024} or the use of GaN devices \cite{8094272}, have pushed efficiency to 98\%. However, most of these works rely on offline multi-variable optimization to generate lookup tables, which increases complexity and reduces control flexibility.

In addition to $Topology\text{-}3$ and $Topology\text{-}4$, a single-stage method known as the isolated Y-rectifier (i-YR) was proposed in \cite{10131553,10706899}. This method employs three-phase Y-configured bridges on the AC side and a three-phase transformer with Y-connected windings. It effectively reduces the total device count and eliminates the need for bidirectional switches present in $Topology\text{-}3$ and $Topology\text{-}4$. However, it suffers from relatively high circulating currents near the grid voltage zero-crossing, and a double switching frequency problem.
\begin{figure*}[!t]
    \centering
    \vspace{-10pt} 
    \includegraphics[width=0.99\linewidth]{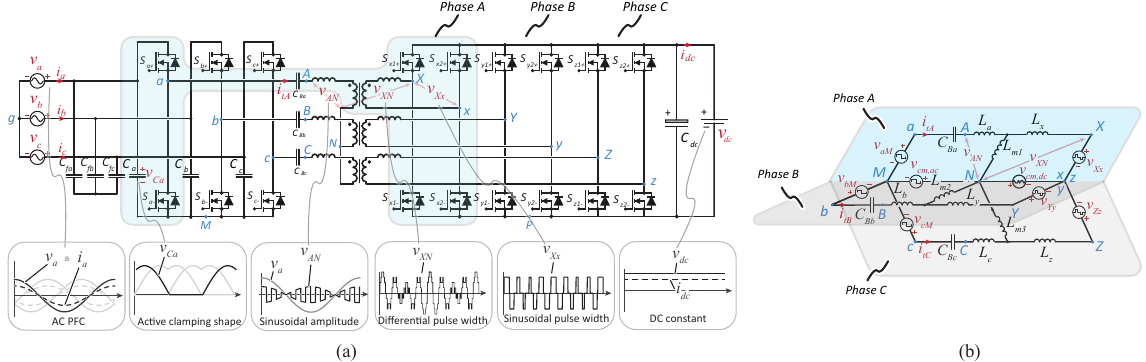}
    \caption{\textbf{(a)} Proposed YAB converter; \textbf{(b)} Equivalent circuit of the YAB converter.}
    \vspace{-10pt} 
    \label{fig:topology2}
\end{figure*}

This paper presents a single-stage YAB converter. By connecting the primary winding of AC-DC DAB ($Topology\text{-}4$) together in a Y-configuration and utilizing the same AC front of the i-YR, the proposed design effectively resolves the control complexity issues commonly found in single-stage converters while achieving high efficiency. Key advantages of this topology include:
\begin{itemize}
    \item Single-stage isolated AC-DC PFC power conversion.
    \item Removal of bulky electrolytic capacitors and grid-side boost inductors.
    \item Elimination of bidirectional switches or unfolding bridges on the AC side.
    \item Straightforward single-variable open-loop control.
    \item Capability to provide a variable DC output voltage.
    \item Soft-switching performance.
    \item High efficiency across a wide operating range.
    \item Reliable and easy startup without soft-start circuits.
\end{itemize}

This paper is organized as follows: Section.\ref{sect:topology} introduces the proposed topology and details the modulation methods. Section.\ref{sec:modeling} derives the numerical steady-state model and analyzes system characteristics including current, power, total harmonic distortion (THD), soft switching performance and device losses. Section.\ref{Sect:hardware} presents the hardware test results, and finally, Section V provides the conclusion and discusses future work. 

\section{Topology and modulation method}\label{sect:topology}
\subsection{Introduction to the Topology}\label{subsect:intro_topo}

The proposed single-stage YAB converter is depicted in Fig.\ref{fig:topology2}(a). The AC side comprises a three-phase Y-configured half-bridge structure identical to i-YR, and the DC side consists of three full bridges with their DC links connected in parallel. The high-frequency transformer (HFT) in between has a Y-connected primary winding, which is the key feature that makes a difference between the YAB and the AC-DC DAB ($Topology\text{-}4$). Blocking capacitors $C_{Ba}$, $C_{Bb}$, and $C_{Bc}$ are placed between the AC-side half-bridges and the HFT to block the low-frequency (LF) component. For clarity, the AC-side half-bridges are denoted as HB-$a$, HB-$b$, and HB-$c$, while the DC-side half-bridges are referred to as HB-$x1$, HB-$x2$, HB-$y1$, HB-$y2$, HB-$z1$, and HB-$z2$.

The three-phase high-frequency (HF) equivalent circuit of the YAB is shown in Fig.\ref{fig:topology2}(b). Considering phase A (highlighted in blue) as an example, the AC-side HF voltage source is the voltage on the switch $S_{a-}$, while the DC-side HF voltage source is the voltage across the switching nodes $X$ and $x$. The transformer leakage inductances on the AC and DC sides are represented by $L_a$ and $L_x$, respectively, and $L_m$ represents the magnetizing inductance. Phases A, B, and C are coupled through nodes $M$, $N$, and ($xyz$).

\subsection{Modulation Method}\label{sect:modulation}
The proposed sinusoidal phase shift (Sin-PS) modulation scheme for the YAB converter operates as follows: the AC side continuously generates bipolar switching pulses with a 50\% duty cycle, while the DC side generates switching pulses with sinusoidal duty cycles ($d_x, d_y, d_z$) corresponding to the grid voltages ($v_a, v_b, v_c$). The main phase shift $\varphi$ between the AC-side and DC-side pulses controls the total power. Since ($d_x, d_y, d_z$) are directly dependent on the grid voltages, the only control variable is the main phase shift $\varphi$, which is identical across all three phases.

\subsubsection{AC Side Modulation}\label{subsect:modulation_type2}
The AC-side Y-configured half-bridges were detailed in \cite{10131553}. The principle is that by disconnecting the grid neutral ($g$) and the switching bridge neutral ($M$), two three-phase three-wire (TPTW) networks are formed: one between $g$ and $N$, and another between $N$ and $M$. Within these networks, any common-mode (CM) voltage does not influence the HFT winding voltage, as it does not generate current through the inductance.

\begin{figure}[!t]
    \centering
    \includegraphics[width=0.9\linewidth]{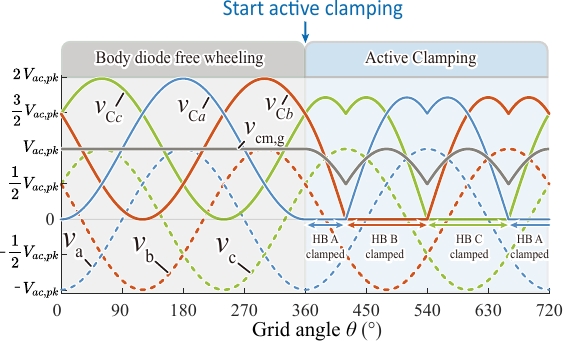}
    \caption{AC side voltages at transition to active clamping. $v_{a},v_{b},v_{c}$ - the grid phase voltages; $v_{Ca},v_{Cb},v_{Cc}$ - the input capacitor voltages; $v_{cm,g}$ the CM voltage in TPTL network between $g$ and $M$.}
    \label{fig:Start-Clamping}
\end{figure}

The body diodes of the three half-bridges naturally force a DC component $v_{cm,g}$ between $g$ and $M$, ensuring the voltages on $C_a$, $C_b$, and $C_c$ remain positive, as shown in the initial stage in Fig.\ref{fig:Start-Clamping}. This feature allows the AC side to use standard unidirectional MOSFETs rather than the bidirectional switches required in $Topology\text{-}3$ and $Topology\text{-}4$. However, as a trade-off, higher voltage-rated MOSFETs are necessary, as the capacitor peak voltage is twice the peak phase voltage. For a 480 V grid, the phase voltage peak is $\hat{v}_g = 391\,\mathrm{V}$, requiring MOSFETs with a blocking voltage exceeding 800 V.

To further reduce device voltage stress and switching losses, the active clamping method proposed in \cite{10438890} is adopted. In this method, each phase is clamped periodically for 120\textdegree, leading to the second stage waveform shown in Fig.\ref{fig:Start-Clamping}. Here, the capacitor voltages $v_{Ca}$, $v_{Cb}$, and $v_{Cc}$ become a camel-hump shape, but the gird phase voltages, as differential-mode (DM) components, retain to be sinusoidal.

\begin{figure}[!t]
    \centering
    \includegraphics[width=0.75\linewidth]{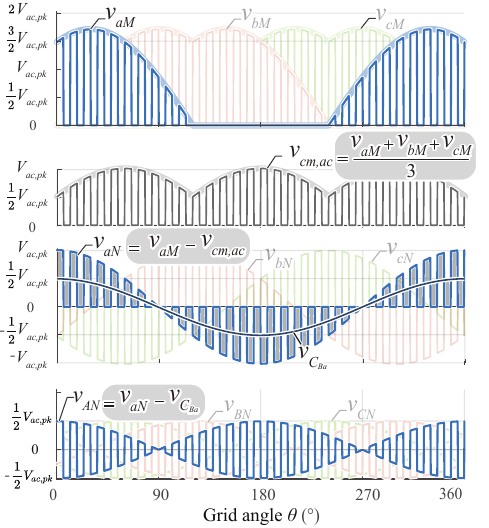}
    \caption{AC-side switching waveforms \cite{10438890}. From top to bottom: $v_{aM},v_{aM},v_{aM}$ - AC side switching node voltages with respect to $M$; $v_{cm,ac}$ - CM voltage in TPTL network between $M$ and $N$; $v_{aN},v_{bN},v_{cN}$ switching node voltages with respect to $N$; and $v_{AN},v_{BN},v_{CN}$ - the HFT primary winding voltages.}
    \label{fig:AC_side_modulation}
\end{figure}

A detailed waveform under this modulation is shown in Fig.\ref{fig:AC_side_modulation}, illustrating the derivation of winding voltage $v_{AN}$.

Considering the loop $M$-a-A-N and applying Kirchhoff's Voltage Law (KVL):
\begin{equation}
\small
    v_{AN} = v_{aM} - v_{C_{Ba}} - v_{cm,ac}.
\end{equation}
Here, $v_{cm,ac}$ the CM voltage in TPTL network between $g$ and $M$:
\begin{equation}
\small
    v_{cm,ac} = \frac{v_{aM} + v_{bM} + v_{cM}}{3}.
\end{equation}
The term $v_{C_{Ba}}$ represents the 60 Hz component, which is fully blocked by $C_{Ba}$:
\begin{equation}
\small
    v_{C_{Ba}} = \frac{1}{2}v_{a}.
\end{equation}

The blocking capacitance $C_B$ must be selected carefully to ensure the LF voltage across the HFT is negligible while avoiding resonance in the circuit. Based on the calculations detailed in Appendix~\ref{appendix:blockingcap_select}, the allowable range for the blocking capacitance is $3.28\,\mu\mathrm{F} < C_B < 86.4\,\mu\mathrm{F}$. For the final design, $C_B = 4.5\,\mu\mathrm{F}$ is selected.

\subsubsection{DC Side Modulation}

\begin{figure}[!t]
    \centering
    \includegraphics[width=0.75\linewidth]{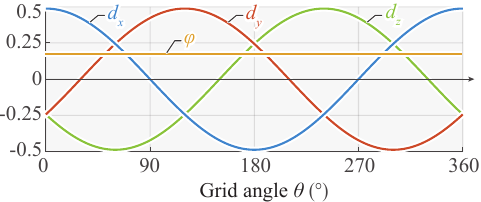}
    \caption{DC side modulation signal}
    \label{fig:5_modulation_signal}
\end{figure}

As illustrated in Fig.\ref{fig:5_modulation_signal}, the DC side employs a simple sinusoidal modulation based on the grid voltage. Assuming the grid voltages are:
\begin{equation}\label{eq:grid_voltage}
\small
    \begin{split}
        v_a &= V_{g} \cdot \cos\left(\frac{\pi}{180}\theta\right),\\
        v_b &= V_{g} \cdot \cos\left(\frac{\pi}{180}\theta - \frac{2\pi}{3}\right),\\
        v_c &= V_{g} \cdot \cos\left(\frac{\pi}{180}\theta + \frac{2\pi}{3}\right),
    \end{split}
\end{equation}
the DC-side pulse widths are determined by:
\begin{equation}\label{eq:dc_pulse_widths}
\small
    \begin{split}
        d_x &= \frac{v_a}{2v_{dc}} \cdot \frac{T_{sw}}{2},\\
        d_y &= \frac{v_b}{2v_{dc}} \cdot \frac{T_{sw}}{2},\\
        d_z &= \frac{v_c}{2v_{dc}} \cdot \frac{T_{sw}}{2},
    \end{split}
\end{equation}
which is achieved through the inner phase shift of the DC-side full bridges.

\begin{figure}[!t]
    \centering
    \includegraphics[width=0.95\linewidth]{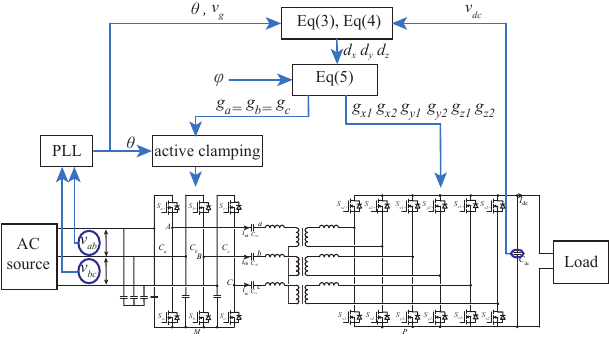}
    \caption{The Sin-PS modulation scheme for the proposed YAB converter.}
    \label{fig:Test_setup1}
\end{figure}

Using the AC-side gate signal (top switch) as a reference, all the DC-side gate signals are phase-shifted by $\varphi$ so that the DC side winding voltage have a phase shift compared to the AC side to make inductive power flow, the same with DAB. Accordingly, the phase shifts of all the gate signals are obtained.
\begin{equation}
\small
\begin{split}
    &ps_{a} = ps_{b} = ps_{c} = 0,\\
    &ps_{x1} = \varphi + \frac{1-d_x}{4}, \quad
    ps_{x2} = \varphi + \frac{1+d_x}{4},\\
    &ps_{y1} = \varphi + \frac{1-d_y}{4}, \quad
    ps_{y2} = \varphi + \frac{1+d_y}{4},\\
    &ps_{z1} = \varphi + \frac{1-d_z}{4}, \quad
    ps_{z2} = \varphi + \frac{1+d_z}{4}.
\end{split}
\end{equation}

This Sin-PS modulation method is illustrated in the flowchart shown in Fig.\ref{fig:Test_setup1}. In this scheme, $d_x$, $d_y$, and $d_z$ are dependent variables generated using a phase-locked loop (PLL) to ensure PFC. The main phase shift $\varphi$ serves as the sole controlled variable to regulate the total system power. The relationship between $\varphi$ and the system power $P$ is analyzed in Sect.\ref{subsect:power_charact}.

Overall, compared to the modulation methods\cite{8094272,6671445,7377097,9583867,9796024,9356460,9351779} for traditional AC-DC DAB ($Topology\text{-}4$), this Sin-PS control scheme is significantly simpler, offering reduced implementation complexity.

\section{Numerical modeling and analysis}\label{sec:modeling}
This section derives the steady state models of the proposed YAB to evaluate its power characteristics, THD, soft-switching performance, MOSFET losses, and magnetic components. Considering phase symmetry, only phase A is considered in the following calculation. To avoid complex switching modes analysis, a universal FFT-based numerical DAB model proposed in \cite{10362268} is adopted. 

In this numerical model, a grid period is sampled into 360 points according to grid angle $\theta$. For each grid angle, the system is approximated as a DC-DC DAB. Therefore, it is assumed that for a given $\theta$, the grid voltage ($v_a$) and blocking capacitor voltage ($v_{C_{Ba}}$) remain constant.

\begin{figure}[!t]
    \centering
    \includegraphics[width=1\linewidth]{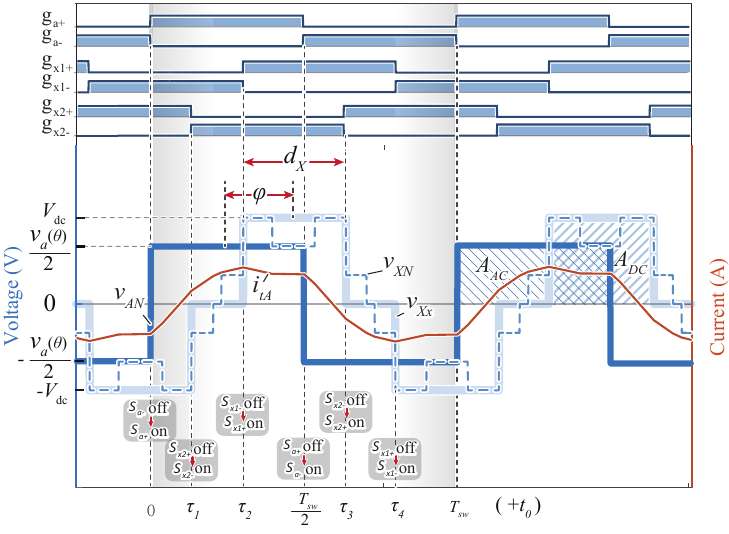}
    \caption{Phase shift modulation of Phase A at grid angle $\theta = 30^\circ$. The AC side winding voltage $v_{AN}$ is a constant 50\% duty cycle square pulse, with an amplitude following grid voltage $\frac{v_a(\theta)}{2}$. The DC side winding voltage $v_{Xx}$ is a three-voltage-level waveform with a constant amplitude of $v_{dc}$, while its duty cycle is sinusoidal modulated according to the principle $A_{DC} = A_{AC}$. $v_{XN}$ is the differential component in $v_{Xx}$ that work with $v_{AN}$ in affecting the winding current.}
    \label{fig:Switching_example}
\end{figure}

A switching period at a given $\theta$ is shown as the grey-shaded area in Fig.\ref{fig:Switching_example} starting from $0$ and ending at $T_{sw}$, All waveforms in this switching period are sampled into arrays of length $N_{sw}$ ($N_{sw}$ must be even).

The derivation starts with obtaining all gate signals $\bm{g}_{a_+}$, $\bm{g}_{x1_+}$, and $\bm{g}_{x2_+}$. Using $\bm{g}_{a+}$ as the reference:
\begin{equation}\label{eq:gate_signal1}
\small
    \bm{g_{a_+}} = [\underbrace{1, 1, \ldots, 1}_{\frac{N_{sw}}{2}}, \underbrace{0, 0, \ldots, 0}_{\frac{N_{sw}}{2}}];
\end{equation}
The others can be obtained by circularly shifting $\bm{g_{a_+}}$:
\begin{equation*}
\small
    \text{for} \quad n \in \{1, 2, \ldots, N_{sw}\},
\end{equation*}
\begin{equation}\label{eq:gate_signal2}
\small
\left\{
\begin{split}
\small
    &\tau_1 = \frac{T_{sw}}{4} + \varphi - \frac{d_x}{2}, \\
    &k_1 = \mod \left(\lfloor n - N_{sw} \frac{\tau_1}{T_{sw}} \rfloor, N_{sw}\right), \\
    &\bm{g_{x1_+}}(n) = \bm{g_{a_+}}(k_1)
\end{split}
\right.
\end{equation}

\begin{equation}\label{eq:gate_signal3}
\small
\left\{
\begin{split}
    &\tau_2 = \frac{T_{sw}}{4} + \varphi + \frac{d_x}{2}, \\
    &k_2 = \mod \left(\lfloor n - N_{sw} \frac{\tau_2}{T_{sw}} \rfloor, N_{sw}\right), \\
    &\bm{g_{x2_+}}(n) = \bm{g_{a_+}}(k_2)
\end{split}
\right.
\end{equation}

\subsection{HFT winding voltage}

The primary and secondary winding voltages are expressed as:
\begin{equation}\label{eq:array_vAN}
\small
    \bm{v_{AN}} = \frac{v_a(\theta)}{2} \cdot \bm{g_{a_+}},
\end{equation}
\begin{equation}\label{eq:array_vXN}
\small
    \bm{v_{Xx}} = v_{dc} \cdot (\bm{g_{x1_+}} - \bm{g_{x2_+}}).
\end{equation}

In traditional AC-DC DAB converters, the difference between the primary and secondary winding voltages directly induces the transformer winding current. However, this is not the case for the YAB, where only the DM voltage contributes to current flow through the transformer due to the Y-connected winding configuration. The CM component induces only a negligible magnetizing current on the DC side.

The DM component $\bm{v}_{XN}$ is obtained by subtracting the CM component $\bm{v}_{cm,dc}$ from $\bm{v}_{Xx}$. 
\begin{equation}\label{eq:Effect_DCside_Pulse}
\small
    \bm{v}_{XN} = \bm{v}_{Xx} - \bm{v}_{cm,dc},
\end{equation}
where the CM component is given by:
\begin{equation}\label{eq:vcm_dc}
\small
    \bm{v}_{cm,dc} = \frac{\bm{v}_{Xx} + \bm{v}_{Yy} + \bm{v}_{Zz}}{3},
\end{equation}

\begin{figure}[!t]
    \centering
    \includegraphics[width=0.7\linewidth]{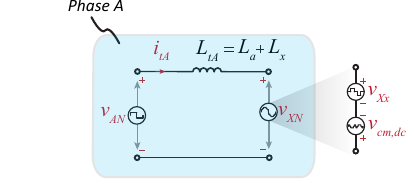}
    \caption{Decoupled single-phase equivalent circuit of phase A.}
    \label{fig:equivalent-circuit2}
\end{figure}

Using the derived AC-side DM pulses ($v_{AN}$, $v_{BN}$, $v_{CN}$) and DC-side DM pulses ($v_{XN}$, $v_{YN}$, $v_{ZN}$), the three-phase equivalent model depicted in Fig.\ref{fig:topology2}(b) can be decoupled into single-phase models. Assuming the transformer magnetizing inductance is much larger than the leakage inductance, the equivalent circuit of phase A finally simplifies to Fig.\ref{fig:equivalent-circuit2}, which is analogous to that of a DC-DC DAB converter.

\subsection{HFT Winding Current}

The voltage applied across the inductance $L_{tA}$ is given by:
\begin{equation}
\small
    \bm{v}_{LA} = \bm{v_{AN}} - \bm{v_{XN}}.
\end{equation}

Using the FFT-based model detailed in \cite{10362268} and the publicly available code on GitHub \cite{mafuGit2024}, the HFT winding current $\bm{i}_{tA}$ for one switching period is computed.

By repeating this calculation for $\theta \in \left[1, 2, \ldots, 360\right]$, a complete inductor current waveform over an entire grid period is obtained. In the MATLAB implementation, this process is vectorized for computational efficiency. The main computational steps can be implemented in a few dozen lines of code, significantly reducing complexity compared to traditional time-domain methods \cite{5776689, 6671445}.

\subsection{Power characteristics}\label{subsect:power_charact}
The grid phase power for phase A at a grid angle $\theta$ is calculated by averaging the instantaneous power over one switching period:
\begin{equation}
\small
    p_a(\theta) = \frac{\sum^{N_{sw}}_{n=1} \bm{v_{AN}}(n) \bm{i}_{tA}(n)}{N_{sw}},
\end{equation} 

\begin{figure}[!t]
    \centering
    \includegraphics[width=0.95\linewidth]{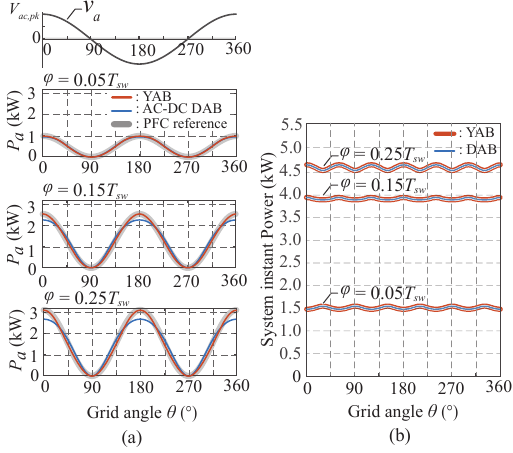}
    \caption{Comparison of power characteristics for the YAB and AC-DC DAB under Sin-PS modulation with $v_{dc} = 200\,\text{V}$. \textbf{(a)} Phase power ($p_a$). \textbf{(b)} Total power ($p$).}
    \label{fig:phase_power}
\end{figure}

To compare the power characteristics of the YAB and the AC-DC DAB, calculations were performed for both using the same Sin-PS modulation and system parameters. The results are presented in Fig.\ref{fig:phase_power}(a). It is observed that the YAB consistently tracks the PFC reference, whereas the AC-DC DAB exhibits deviations, particularly at higher power levels.

The total power $p(\theta)$, shown in Fig.\ref{fig:phase_power}(b), is the summation of three phases. The average total power $P$ can be obtained by averaging $p(\theta)$ over a grid period.

A minor ripple is observed in $p(\theta)$, which is caused by a slight distortion within the limits defined by the IEEE 1547 standard. Interestingly, although the phase powers of the YAB and AC-DC DAB differ under Sin-PS modulation, their total power $p(\theta)$ is always identical. As a result, the closed-form power equation for the YAB is the same as that for the AC-DC DAB, as derived in \cite{9583867}.

\begin{figure}[!t]
    \centering
    \includegraphics[width=0.75\linewidth]{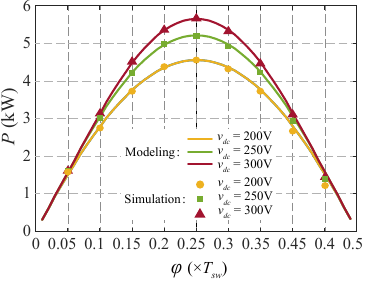}
    \caption{Relationship between system total power $P$ and $\varphi$ at different DC output voltages.}
    \label{fig:power_vs_PS}
\end{figure}

The relationship between the average total power $P$ and the phase shift $\varphi$ was calculated and verified through PLECS simulations. The results are presented in Fig.\ref{fig:power_vs_PS}. It is observed that as the control variable $\varphi$ increases, the system power first rises, reaches its peak at $\varphi = 0.25T_{sw}$, and then begins to decrease. Additionally, higher DC-side voltages result in greater output power, demonstrating power characteristics similar to those of a DC-DC DAB converter.

Although the system power is symmetric in the ranges $\varphi \in [0,0.25]T_{sw}$ and $\varphi \in (0.25,0.5]T_{sw}$, the system operation is restricted to $\varphi \in [0,0.25]T_{sw}$ to minimize device losses. This will be further analyzed in the MOSFET losses presented in Sect.\ref{Subsect:loss_model}.

\subsection{Current characteristics} 
\subsubsection{Total harmonic distortion}

As observed in Fig.\ref{fig:phase_power}, the YAB outperforms the AC-DC DAB in tracking the PFC under Sin-PS modulation. Deviation from the PFC can lead to increased total harmonic distortion (THD) in the grid current. Therefore, it is essential to analyze the current THD under varying power levels and output voltages and compare it between the YAB and the AC-DC DAB.

As a DAB-type converter with open-loop control, the AC phase voltage $v_a$ is fixed by the grid. The grid current can be calculated as:
\begin{equation}
    i_a(\theta) = \frac{p_a(\theta)}{v_a(\theta)}.
\end{equation}
Using the FFT in MATLAB, the harmonic components' magnitudes $I_{a1}, I_{a2}, \ldots, I_{an}$ of $\bm{i_a}$ are obtained. The total harmonic distortion (THD) is then computed as:
\begin{equation}
    THD = \frac{\sqrt{I_{a2}^2 + I_{a3}^2 + \cdots + I_{an}^2}}{I_{a1}}.
\end{equation}

\begin{figure}[!t]
    \centering
    \includegraphics[width=0.95\linewidth]{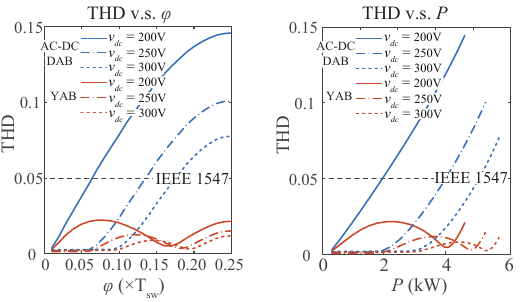}
    \caption{Comparison of current THD for the YAB and AC-DC DAB under Sin-PS modulation within the allowable $\varphi$ range.}
    \label{fig:THD}
\end{figure}

The results are calculated for $\varphi \in [0, 0.25]T_{sw}$ and are presented in Fig.\ref{fig:THD}. Within the operating range of $\varphi$ and $v_{dc}$, the YAB maintains a current THD below 2.5\%, which complies with the IEEE-1547 standard. In contrast, the AC-DC DAB exhibits increasing THD as power levels rise, exceeding the IEEE-1547 limit at higher power levels. This explains why the AC-DC DAB often requires more complex control strategies - even if using the simplest DC-side phase-shift modulation with a constant $\varphi$, online compensation for $d_x$, $d_y$, and $d_z$ is necessary, as discussed in \cite{8643038,10285428}.

\subsubsection{System current stress} 
In this paper, we define the RMS value of the HFT winding current as the system current stress, as it directly impacts the MOSFET conduction loss and HFT winding loss. This value can be easily calculated in MATLAB as:
\begin{equation} 
    I_{tA,\text{rms}} = \text{RMS}([\bm{i}_{tA}|_{\theta=1}, \bm{i}_{tA}|_{\theta=2}, \ldots, \bm{i}_{tA}|_{\theta=360}]).
\end{equation}

\begin{figure}[!t]
    \centering
    \includegraphics[width=0.85\linewidth]{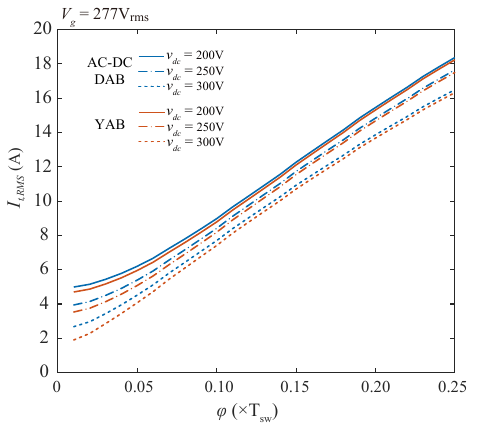}
    \caption{Comparison of current stress for the proposed YAB and the AC-DC DAB under different $\varphi$ and DC output voltages.}
    \label{fig:current_stress}
\end{figure}

The results for current stress are compared between the YAB and AC-DC DAB under various $\varphi$ and $v_{dc}$ conditions, as shown in Fig.\ref{fig:current_stress}. It is evident that for the same power (since the same $\varphi$ yields the same $P$, as proven in Sect.\ref{subsect:power_charact}), the current stress of the YAB is consistently lower than that of the AC-DC DAB across all operating points.

These results for THD and current stress clearly demonstrate that the proposed YAB outperforms the AC-DC DAB under Sin-PS modulation. In the following subsection, we will show that the YAB also require smaller magnetic components than the AC-DC DAB.

\subsection{Magnetic consideration}

In this subsection, the magnetic components of the YAB and the AC-DC DAB are compared. The extreme condition is considered where both the leakage inductance and the extra inductance are on one side of the transformer (e.g., the secondary side).

For the HFT, the maximum flux density occurs during system startup because, at this time, the DC-side voltage is zero, and the primary winding voltage $v_{AN}$ solely magnetizes the transformer. Since $v_{AN}$ is identical for the YAB and AC-DC DAB, the same core size should be selected for both topologies.

However, the power capability of the YAB and the AC-DC DAB differs with identical parameter designs. As illustrated in Fig.\ref{fig:phase_power}(a), the maximum phase power that the AC-DC DAB can achieve with a fixed $\varphi$ is consistently lower than the YAB. Therefore, after applying compensation on $d_x$, $d_y$, and $d_z$ to reduce THD, as discussed in \cite{8643038,10285428}, the AC-DC DAB total power becomes constantly lower than the YAB. Due to this, it is more appropriate to compare a normalized value $\Bar{B}_{max} = B_{a,\text{max}} / \hat{p}_a$, which represents the ratio of peak flux density to peak power of one phase.

\begin{figure}[!t]
    \centering
    \includegraphics[width=0.95\linewidth]{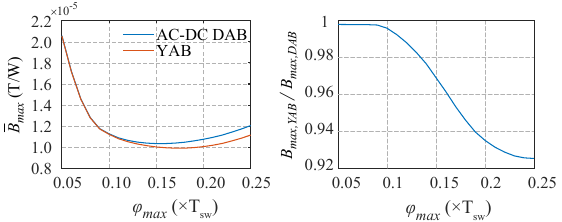}
    \caption{Normalized inductor peak flux density for the YAB with Sin-PS modulation and the AC-DC DAB with DC-side compensated modulation. Operating conditions: $v_g = 277\,\mathrm{V}_{\mathrm{rms}}$, $v_{dc} = 300\,\mathrm{V}$.}
    \label{fig:B_max_nom}
\end{figure}

While the HFT size is fixed due to the startup condition, adjustments can be made to the inductor design. Detailed calculations for $\Bar{B}_{max}$ in the inductors are provided in Appendix~\ref{appendix:Magnetic}, and the results are shown in Fig.\ref{fig:B_max_nom}. It is evident that the normalized inductor flux density for the YAB is consistently lower than that of the AC-DC DAB. This indicates that smaller magnetic components can be used in the YAB for the same power rating.

In an optimized hardware design, an operational range for $\varphi \in [\varphi_{\text{min}}, \varphi_{\text{max}}] \subseteq [0, 0.25]T_{sw}$ should be selected, and the hardware should be designed accordingly to ensure high power density and efficiency, similar to DAB designs. The upper bound $\varphi_{\text{max}}$, which corresponds to the maximum power, is typically set below $0.25T_{sw}$ to achieve higher overall efficiency by reducing circulating current. However, Fig.\ref{fig:B_max_nom} suggests that $\varphi_{\text{max}}$ should not be too low, as values below $0.1T_{sw}$ result in high $\Bar{B}_{\text{max,DAB}}$, which reduces the power density of the magnetic components. This trade-off should be carefully considered during hardware design.

For this study, all magnetic components were intentionally designed to exceed the size requirements for $\varphi_{\text{max}} = 0.25T_{sw}$. This approach ensures that the topology principles are validated without introducing constraints from hardware limitations.

It is important to note that the extra inductors are not mandatory components for this topology. They were added in this study to limit system power. Alternatively, these inductors can be eliminated by designing a transformer with the required leakage inductance.

\subsection{Solf switching performance}\label{Subsect:ZVS}
The criterion for determining ZVS is that the device current must be negative at the instant of turning on, i.e., the current must be flowing through the body diode so that the turn-on voltage is zero. According to Fig.\ref{fig:Switching_example}, the turn-on current of each device at a given grid angle $\theta$ can be obtained as follows:

- $t=0$: $S_{a+}$ turns on. $i_{sw,a+}(\theta) = \bm{i}_{tA}(1)|_{\theta\in[1,360]}$.

- $t=\tau_2$: $S_{x1+}$ turns on. $i_{sw,x1+}(\theta) = -\bm{i}_{tA}(\lfloor\frac{\tau_2}{T_{sw}}N_{s}\rfloor)|_{\theta\in[1,360]}$.

- $t=\tau_3$: $S_{x2+}$ turns on. $i_{sw,x2+}(\theta) = \bm{i}_{tA}(\lfloor\frac{\tau_3}{T_{sw}}N_{s}\rfloor)|_{\theta\in[1,360]}$.

\begin{figure}[!t]
    \centering
    \includegraphics[width=1\linewidth]{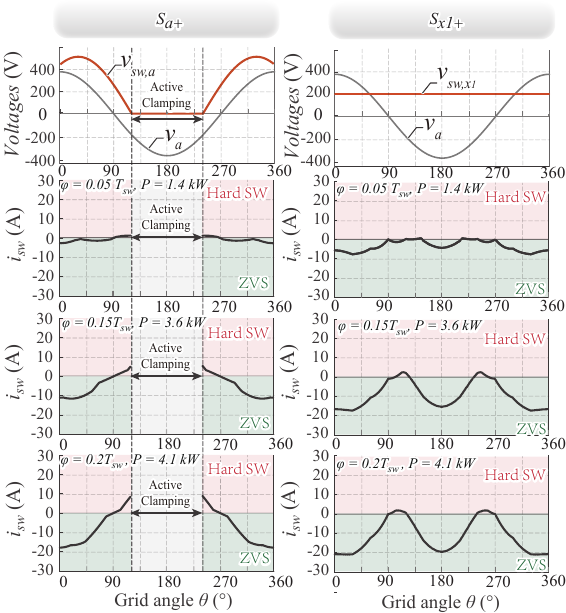}
    \caption{Turn on current of $S_{a+}$ and $S_{x1+}$ at different power level, i.e., different $\varphi$. Calculated operating point: $v_g=277V_{rms}$, $v_{dc}=200V$}
    \label{fig:ZVS_mainfig}
\end{figure}

Due to symmetry, the switching status of the bottom switch is the same as the top switch. Therefore, only the top switches need to be investigated. Moreover, due to the phase shift modulation, turn-on currents of HB-$x1$ and HB-$x2$ are symmetrical within the grid period. Therefore, finally only $S_{a+}$ and $S_{x1+}$ need to be investigated. The results are shown in Fig.\ref{fig:ZVS_mainfig}, with $v_{dc}=200V$.

The left are the results for $S_{a+}$. the switching voltage $v_{sw,a}$ is shown in the first plot, which is calculated with
\begin{equation}
    v_{sw,a}(\theta) =
\begin{cases} 
\sqrt{6}\frac{v_g}{v_{dc}}sin(\frac{2\pi}{3}-\frac{\theta}{180}\pi) & \text{if } \theta \leq 120, \\
\sqrt{6}\frac{v_g}{v_{dc}}sin(\frac{2\pi}{3}+\frac{\theta}{180}\pi) & \text{if } \theta \geq 240.\\
0 &\text{if } 120 < \theta < 240
\end{cases}
\end{equation}

From $0^\circ$ to $90^\circ$, the turn-on current $i_{sw,a}$ is always negative across all power levels, indicating that $S_{a+}$ achieves ZVS throughout this period across all power level. Between $90^\circ$ and $120^\circ$, $i_{sw,a}$ becomes positive and results in hard switching, however, this interval is very short, and the switching voltage $v_{sw,a}$ is decreasing to zero when $i_{sw}$ increases. Consequently, the switching loss during this period is low. During $120^\circ < \theta < 240^\circ$, which is 1/3 of the entire period, there is no loss since the voltage on HB-$a$ is clamped to zero.

The results of $S_{x1+}$ is shown on the right in Fig.\ref{fig:ZVS_mainfig}. Notably, the switching voltage always equals $v_{dc}$.
For the majority of the grid period, $S_{x1+}$ achieves ZVS across different power levels. Brief intervals of ZCS occur near the grid zero-crossing points.

\subsection{MOSFET Loss} \label{Subsect:loss_model}

For simplicity, switching losses and conduction losses are evaluated at the half-bridge level rather than for individual switches. The detailed calculation is provided in Appendix~\ref{appendix:mosfet_loss}.  

\begin{figure}[!t]
    \centering
    \includegraphics[width=0.9\linewidth]{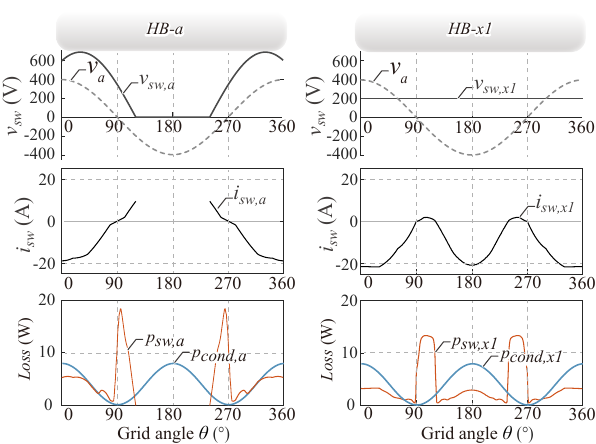}
    \caption{Switching and conduction losses of a half-bridge over a grid period. The operating conditions are: $v_g = 277\,\mathrm{V}_{\mathrm{rms}}$, $v_{dc} = 200\,\mathrm{V}$, $\varphi = 0.2T_{sw}$, and $P = 4.18\,\mathrm{kW}$.}
    \label{fig:mosfet_loss}
\end{figure}

The results in Fig.\ref{fig:mosfet_loss} illustrate the loss distribution over a grid period for the AC-side and DC-side half-bridges. For the AC-side switch ($S_{a+}$), predominantly ZVS operation and the switching-loss-free active clamping section result in low overall switching losses.

On the DC side, ZVS is also achieved for most of the grid period. The ZCS intervals, while associated with relatively higher losses, are limited to very short durations. As a result, the overall switching loss on the DC side remains low.

Conduction losses, being proportional to $I_{tA,\text{rms}}^2$, peak at the grid voltage maxima. Conversely, switching losses peak at the grid voltage zero-crossings. This leads to a relatively even distribution of total heat generation across the grid period, which simplifies thermal design.

\begin{figure}[!t]
    \centering
    \includegraphics[width=0.85\linewidth]{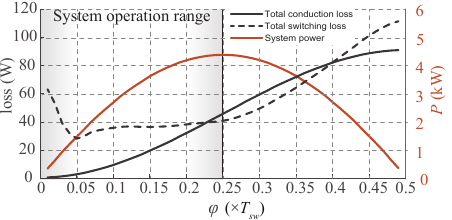}
    \caption{Total MOSFET conduction and switching losses, and system power versus $\varphi$, with $v_{dc} = 200\,\mathrm{V}$.}
    \label{fig:Efficiency_and_loss}
\end{figure}

The total switching and conduction losses, averaged over a grid period, are calculated for $\varphi \in [0, 0.5]T_{sw}$ with $v_{dc} = 200\,\mathrm{V}$ and are shown in Fig.\ref{fig:Efficiency_and_loss}.

It can be observed that switching losses are higher when the total power $P$ is near zero. Verification using PLECS simulation indicates that soft-switching performance deteriorates when the system's total power is below 15\%. However, this does not compromise hardware reliability, as total device losses remain lower than at peak power, which is the basis for heat sink design. A modified modulation method to enhance soft-switching performance during low-power operation will be addressed in future work.

Fig.\ref{fig:Efficiency_and_loss} also highlights why the control variable $\varphi$ should be constrained to $[0, 0.25]T_{sw}$ - both switching and conduction losses are significantly higher in $[0.25, 0.5]T_{sw}$.

\section{Hardware test}\label{Sect:hardware}
A YAB converter prototype, with the parameters listed in Table. \ref{tab:system_parameters}, was tested in the lab under a grid voltage of 480 V and a load voltage ranging from 200 V to 300 V. The DC side used an adjustable resistive load.
\begin{table}[h]
\centering
\caption{System Parameters}
\begin{tabular}{lll}
\hline
Parameter & Description                 & Value              \\ \hline
$SW$       & Selected MOSFET            & C3M0021120K         \\
$v_{dc}$   & DC voltage                 & 200V$\sim$300V     \\
$v_g/v_{LL}$ & Grid phase/line-line voltage & 277/480V$_{rms}$ \\
$L_\mu$    & HFT leakage inductance     & 4.7 $\mu$H         \\
$L_e$      & Extra inductance           & 14.6 $\mu$H        \\
$L_t = L_\mu + L_e$ & Total inductance      & 19.3 $\mu$H        \\
$C_{Ba}, C_{Bb}, C_{Bc}$ & Blocking capacitance & 4.5 $\mu$F       \\
$C_a, C_b, C_c$ & AC input capacitance   & 0.5 $\mu$F         \\
$C_{fa}, C_{fb}, C_{fc}$ & Grid filter capacitance & 10 $\mu$F      \\
$C_{dc}$ & Output DC-link capacitance   & 10 $\mu$F          \\
$P$  & System total power         & 0kW$\sim$6kW    \\
$f_{sw}$   & Switching frequency        & 100kHz             \\
$A_{c,t}$  & HFT core cross-section area & 7.84$\times$10$^{-4}$ m$^2$ \\
$N_{pri}$  & Primary winding turns      & 21                 \\
$N_{sec}$  & Secondary winding turns    & 21                 \\
$R_{t}$    & HFT resistance (100kHz) & 18 m$\Omega$ \\
$A_{c,l}$  & Inductor core cross-section area & 1.56$\times$10$^{-3}$ m$^2$ \\
$N_{l}$    & Inductor winding turns     & 6                  \\
$R_{l}$    & Inductor resistance (100kHz) & 6.8 m$\Omega$ \\
\hline
\label{tab:system_parameters}
\end{tabular}
\end{table}

\begin{figure}[!t]
    \centering
    \includegraphics[width=0.9\linewidth]{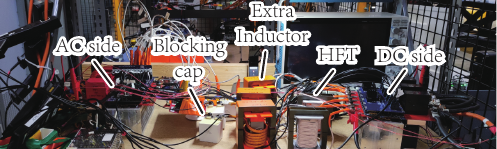}
    \caption{Experimental setup of the proposed YAB converter.}
    \label{fig:Test_setup}
\end{figure}

\subsection{Start up}
\begin{figure}[!t]
    \centering
    \includegraphics[width=0.95\linewidth]{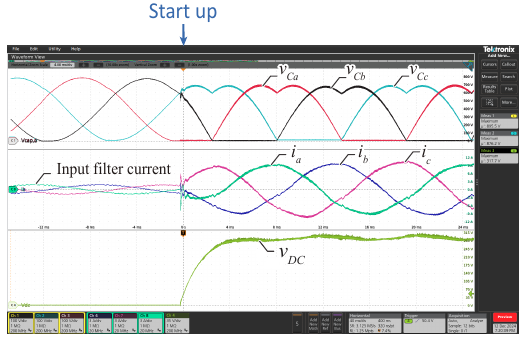}
    \caption{Hardware test result of starting up the YAB converter.}
    \label{fig:hardware_startup}
\end{figure}
Before initialization, the converter was connected to the grid with all switches turned off. As described previously in Sect.\ref{subsect:modulation_type2} and in Fig.\ref{fig:Start-Clamping}, the AC-side CM voltage \(v_{cm,g}\) is naturally established and produce the positive half-bridge input voltages \(v_{Ca}\), \(v_{Cb}\), and \(v_{Cc}\) (see Fig.\ref{fig:hardware_startup}). During this time, there will be some low current on the grid due to the input filter capacitor.

Once \(v_{Ca}\) reaches zero, the proposed Sin-PS modulation with active clamping is initiated, and the system begins delivering power to the DC resistive load. The DC side voltage is gradually increased, until it reach the designed $v_{dc}$. At the starting transient, grid currents experienced small resonance and soon gained their sinusoidal waveform, with no in-rush current observed. The input capacitor voltage is also smoothly transformed into the hump shape without overshot. This result proves the benefit of easy starting up without extra soft-start circuits as mentioned in the introduction.
\begin{figure}[!t]
    \centering
    \includegraphics[width=0.85\linewidth]{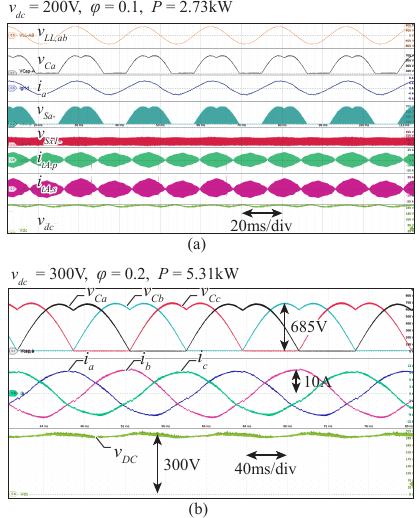}
    \caption{Hardware test result. \textbf{(a)} The waveforms of phase A at $v_{dc}=200$V $P=2.73$kW. From top to bottom, the respective waveforms are: $v_{LL,ab}$ - the grid line-to-line voltage between phase A and phase B; $v_{Ca}$ - the phase A input capacitor voltage; $i_a$ - the grid current; $v_{Sa-}$ - the voltage on switch $S_{a-}$; $v_{Sx1-}$ - the voltage on switch $S_{x1-}$; $i_{tA,p}$ and $i_{tA,s}$ - the leakage inductor currents on the primary and secondary sides; and $v_{dc}$ - the DC output voltage. \textbf{(b)} The three phase waveforms at $v_{dc}=300$V $P=5.31$kW.}
    \label{fig:HW_main}
\end{figure}

\subsection{Steady state results}
The phase A waveform at $v_{dc}=200$ V and $P=2.73$ kW is shown in Fig.\ref{fig:HW_main}(a), whereas the three-phase waveforms under higher voltage and power conditions ($v_{dc}=300$ V and $P=5.31$ kW) are presented in Fig.\ref{fig:HW_main}(b). In Fig.\ref{fig:HW_main}(a), the grid current $i_a$ is observed to slightly lead the input capacitor voltage $v_{Ca}$, i.e. lead the grid phase voltage $v_a$. This occurs because the filter capacitor current shown in the initial time segment of Fig.\ref{fig:hardware_startup} is added to the converter PFC current. Such behavior is common in grid-tied rectifiers with input filters~\cite{6671445}. With higher power, the leading angle diminishes, as shown in Fig.\ref{fig:HW_main}(b). Eliminating the leading current entirely would require a PQ controller, which falls outside the scope of this paper.

The input capacitor voltage exhibits a well-defined hump-shaped waveform, while the grid currents maintain a high-quality sinusoidal shape with low distortion.

\subsection{Switching pulses and soft switching inspection}
\begin{figure}[!t]
    \centering
    \includegraphics[width=0.9
    \linewidth]{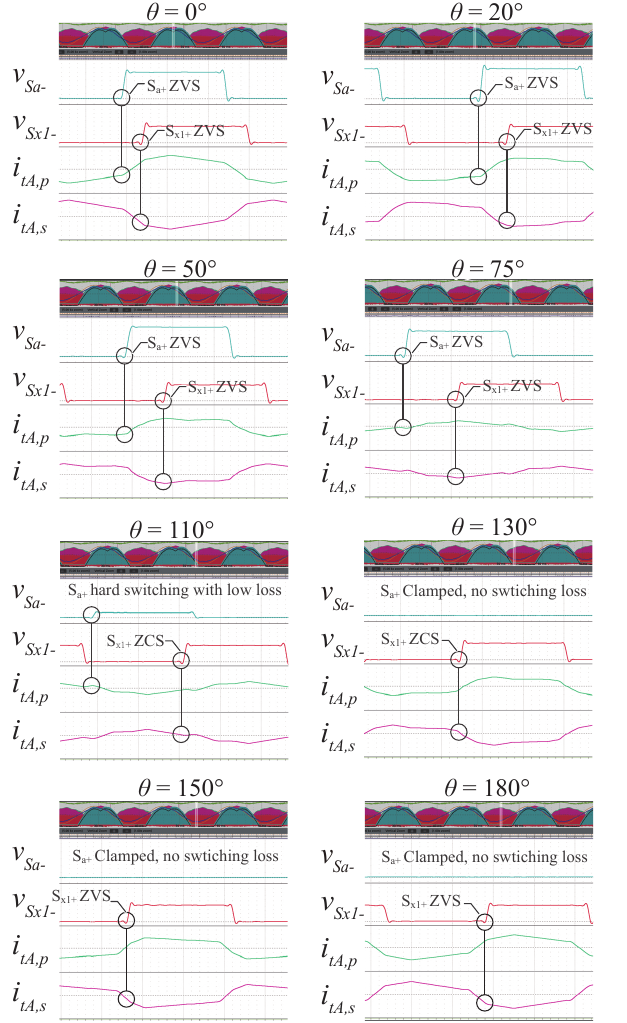}
    \caption{Measured MOSFET voltages and winding currents at various angles from $0^\circ$ to $180^\circ$, illustrating the soft switching status. The operating conditions are $v_{dc} = 200$ V and $P = 2.73$ kW.}
    \label{fig:HW_ZVS}
\end{figure}

As discussed in Sect.\ref{Subsect:ZVS}, due to symmetry, it suffices to only analyze the switching currents of $S_{a+}$ and $S_{x1+}$. In Fig.\ref{fig:HW_ZVS}, the switch voltages $v_{Sa-}$, $v_{Sx1-}$, and the inductor currents $i_{tA,p}$ and $i_{tA,s}$ are magnified to examine the soft switching status. The currents $i_{tA,p}$ and $i_{tA,s}$ are measured in such directions that positive values indicate current flowing out of the half-bridge (HB) switching node. The rising edges of $v_{Sa-}$ and $v_{Sx1-}$ correspond to the turn-on instances of $S_{a+}$ and $S_{x1+}$, respectively. Therefore, if the measured current is negative at these instances, the corresponding half-bridge achieves ZVS.

The results demonstrate that ZVS is achieved for both HB-$a$ and HB-$x1$ at 0\textdegree, 20\textdegree, 50\textdegree, and 70\textdegree. Hard switching occurs at 110\textdegree, but due to the low switching voltage and current, the switching loss remains low. At 130\textdegree 150\textdegree and 180\textdegree, HB-$a$ is clamped, eliminating switching losses during this interval, while HB-$x1$ continues to achieve ZVS.

These test results align closely with the numerical modeling results shown in Fig.\ref{fig:ZVS_mainfig}, validating the accuracy of the numerical model.

\subsection{Power and efficiency}

\begin{figure}[!t]
    \centering
    \includegraphics[width=0.95\linewidth]{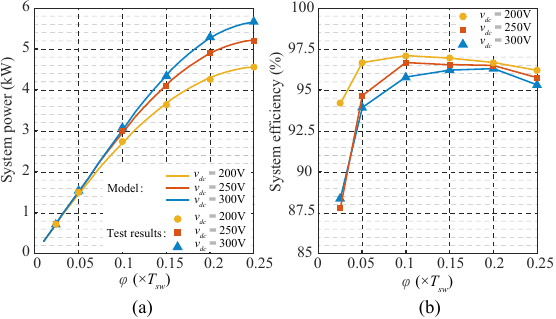}
    \caption{\textbf{(a)}Measured total power $P$ v.s. $\varphi$ and comparison with the numerical model. \textbf{(b)} Measured efficiency v.s. $\varphi$ at different output voltages }
    \label{fig:power_efficiency_test}
\end{figure}

The system power \( P \) was measured across the entire operating range of \(\varphi \in [0, 0.25]T_{sw}\) under different output voltage conditions. The results are presented in Fig.\ref{fig:power_efficiency_test}(a). The measured power aligns closely with the modeled results derived in Sect.\ref{subsect:power_charact}, validating the accuracy of the theoretical predictions.

Fig.\ref{fig:power_efficiency_test}(b) illustrates the measured efficiency across varying output voltages and $\varphi$. The peak efficiency of 97.14\% was observed at \(\varphi = 0.1T_{sw}\) and \(v_{dc} = 200\) V. Within the examined voltage range of 200 V to 300 V, a clear trend emerges: the efficiency decreases as the output voltage increases. This behavior is consistent with the characteristics of AC-DC DAB converters, where the maximum efficiency is typically achieved when the DC-side voltage is equal to the peak AC-side voltage. The higher the boost rate, the lower the efficiency \cite{9583867}. For this converter, which features half-bridges on the AC side and full-bridges on the DC side, the optimal efficiency should occurs when \(\hat{v}_{ac} = 2v_{dc}\), i.e., \(v_{dc} \approx 200\) V, as observed in the experimental results.

\section{Conclusion}
This paper proposes a novel YAB converter capable of single-stage isolated three-phase AC-DC PFC conversion. Compared to multi-stage designs, the YAB eliminates bulky electrolytic capacitor banks and input boost inductors, while also offering an easy start-up process without the inrush current issue. It inherits the advantages of single-stage AC-DC DAB converters but greatly reduces control complexity. The proposed converter achieves excellent soft-switching performance and demonstrates a high efficiency of 97.1\% in lab tests, making it competitive among the same type of converters. Based on a numerical FFT model, this work details the operating principles, equivalent circuit model, and performance of the YAB. Future work will present a full-operation-range ZVS modulation method to further enhance efficiency, along with testing bidirectional power flow capabilities.

\appendices
\section{Blocking capacitor selection}\label{appendix:blockingcap_select}
\subsubsection{Upper limit}The first condition ensures that the grid frequency component is fully blocked to prevent saturation of the high-frequency transformer (HFT). To achieve this, the flux density generated by the grid frequency voltage must be negligible compared to that generated by the high-frequency voltage pulses. This requirement imposes an upper limit on the capacitance, as a smaller blocking capacitor is necessary. The maximum flux density generated by the HF pulse is given by:
\begin{equation}\label{eq:max_flux}
\small
    B_{pk,HF} = \frac{1}{A_c} \frac{v_g}{2N} \times \frac{1}{4} \frac{1}{f_{sw}},
\end{equation}
whereas the maximum flux density generated by the grid frequency voltage is:
\begin{equation}
\small
    B_{pk,LF} = \frac{1}{A_c} \frac{v_g}{2N} \times \frac{2\pi f_g L_t}{2\pi f_g L_t + \frac{1}{2\pi f_g C_B}} \times \frac{1}{4} \frac{1}{f_g}.
\end{equation}
To ensure that the flux density contribution from the grid frequency is negligible in transformer design, the following inequality is applied:
\begin{equation}
\small
    \frac{B_{pk,LF}}{B_{pk,HF}} < \varepsilon,
\end{equation}
where $\varepsilon = 0.01$. By satisfying this condition, $B_{pk,LF}$ can be ignored during transformer design. Consequently, the upper limit of the capacitance is derived as:
\begin{equation}
\small
    C_B < \frac{\varepsilon}{f_g(f_{sw} - \varepsilon f_g)L_t}.
\end{equation}

\subsubsection{Lower limit}The second condition requires that the capacitance value is not too low, ensuring that the resonant frequency of the L-C loop remains higher than the switching frequency:
\begin{equation}
\small
    f_r = \frac{1}{2\pi\sqrt{L_t C_B}},
\end{equation}
with the following constraint:
\begin{equation}
\small
    \frac{f_r}{f_{sw}} < \lambda.
\end{equation}
The lower limit of the capacitance is therefore given by:
\begin{equation}
\small
    C_B > \frac{1}{4\pi^2 f_{sw}^2 L_t}.
\end{equation}

Considering $\lambda = 0.2$, the YAB demonstrates inductive high-frequency power transfer. Although $\lambda$ can be increased up to $0.85$ to operate the L-C tank in a quasi-resonant mode, which improves system efficiency by reducing circulating power and transformer AC resistance as discussed in \cite{9796024}, this work focuses on regular DAB-type operation. Thus, a stricter constraint of $\lambda = 0.2$ is applied.
\section{Magnetic components comparison}\label{appendix:Magnetic}

As discussed in Sect.\ref{sect:topology} For YAB, the inductor voltage within a switching period is
\begin{equation}
\small
    v_{L,{YAB}}=v_{AN}-v_{XN}.
\end{equation}
For AC-DC DAB, it is
\begin{equation}
\small
    v_{L,{DAB}}=v_{AN}-v_{Xx}
\end{equation}
The maximum flux density of them can be obtained as
\begin{equation}
\small
    B_{max,{YAB}} =\frac{1}{4N_{l}N_{sw}f_{sw}A_{c,l}}||\bm{v}_{LA,YAB}|_{\theta=0}||_1
\end{equation}
\begin{equation}
\small
    B_{max,{DAB}} = \frac{1}{4N_{l}N_{sw}f_{sw}A_{c,l}}||\bm{v}_{LA,DAB}|_{\theta=0}||_1
\end{equation}
Since the AC-DC DAB has a current distortion under the Sin-PS modulation, compensation in DC side phase shifts is needed to reduce the THD\cite{8643038}. Due to the fact that the phase peak power that AC-DC DAB can achieve is always lower than the YAB with the same system parameter (Fig.\ref{fig:phase_power}(a)), it is more reasonable to compare the normalized quantity
\begin{equation}
\small
    \Bar{B}_{max,YAB}=\frac{B_{max,{YAB}}}{\hat{p}_{a,YAB}} =\frac{B_{max,{YAB}}}{p_{a,YAB}(0)}
\end{equation}
and
\begin{equation}
\small
    \Bar{B}_{max,DAB}=\frac{B_{max,{DAB}}}{\hat{p}_{a,DAB}}=\frac{B_{max,{DAB}}}{p_{a,DAB}(0)}
\end{equation}

All the parameters in the calculation are listed in Table.\ref{tab:system_parameters}.

\section{MOSFET loss calculation}\label{appendix:mosfet_loss}
\subsection{Switching loss}
In the case of hard switching, the half-bridge switching loss is the sum of the turn-on loss of the turned-on device and the turn-off loss of the complementary switch. Conversely, when ZVS occurs, the half-bridge switching loss consists only of the turn-off loss of the complementary device, because the turn-on loss is zero.

Partial ZVS occurs when $i_{sw}$ is negative but insufficient to fully commutate the $C_{oss}$ of both devices. The condition for complete commutation is expressed as:
\begin{equation}
\small
    i_{sw} > \frac{v_{sw}}{\sqrt{\frac{L_t}{2C_{oss}}}}
\end{equation}

\begin{figure}[h]
    \centering
    \includegraphics[width=0.75\linewidth]{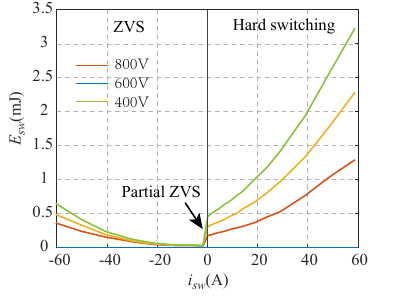}
    \caption{Switching loss map of a half bridge using the selected SiC MOSFET C3M0021120K\cite{Wolfspeed_thermol}, with external gate resistor of $R_g = 10\,\Omega$ and a junction temperature of $T_j = 100^\circ\mathrm{C}$.}
    \label{fig:lossmap}
\end{figure}

A switching loss map (Fig.\ref{fig:lossmap}) for the C3M0021120K is available on \cite{Wolfspeed_thermol}, with which the switching loss of a half bridge can be determined through interpolation:
\begin{equation}
    \small
    p_{sw}(\theta) = f_{sw}E_{sw}(v_{sw}(\theta), i_{sw}(\theta)),
\end{equation}
 \subsection{Conduction loss}
 For a DC-side half-bridge (HB-$x1$ for example), only one switch conducts at any given time. Ignoring switching transients, the conduction loss at a grid angle $\theta$ can be calculated with
\begin{equation}\label{eq:conduction_loss_trains}
\small
    p_{cond,x1}(\theta) = \sqrt{\frac{1}{N_{sw}}||\bm{i}_{tA|_{\theta}}||^2 _2} R_{ds,\text{on}},
\end{equation}
in which the $\bm{i}_{t|_{\theta}}$ is the transformer current waveform vector of phase A at grid angle $\theta$.

For the AC side, the conduction loss is theoretically lower, because both the upper and lower switches conduct during the active clamping interval. In this case, the current splits to two switches, therefore the equivalent resistance is lower than $R_{ds,on}$. However, this effect is ignored here since the lower path has a higher resistance due to the capacitor \( C_a \), which introduces both its internal resistance and an impedance of \( 1 / (2 \pi f_{sw} C_a) \). Consequently, the majority of the current continues to flow through the upper switch, therefore conduction loss is calculated using the same formula as in Eq.~\eqref{eq:conduction_loss_trains}.

\bibliographystyle{IEEEtranTIE}
\bibliography{BIB}\ 
\newpage
\begin{IEEEbiographynophoto}{Mafu Zhang} received the B.S. degree in electrical engineering and automation, from Xi'an Jiaotong University, Xi'an, China, in 2017, the M.Sc. degree in energy science and technology from Power Electronic System (PES) Lab, ETH Z{\"u}rich, Z{\"u}rich, Switzerland, in 2021. He is currently working toward the Ph.D. degree with the Semiconductor Power Electronics Center (SPEC), The University of Texas at Austin, Austin, TX, USA.
\end{IEEEbiographynophoto}
\vspace{-35pt}
\begin{IEEEbiographynophoto}{Huanghaohe Zou} received the B.S. degree in electrical engineering from Xi'an Jiaotong University, Xi'an, China, in 2019. He is currently working toward the Ph.D. degree in electrical and computer engineering with the University of Texas at Austin, Austin,
TX, USA.
\end{IEEEbiographynophoto}
\vspace{-35pt}
\begin{IEEEbiographynophoto}{Saleh Farzamkia} received the B.S. degree (with Hons.) from Bu-Ali
Sina University, Hamedan, Iran, in 2013, and the
M.Sc. degree (with Hons.) from the University of
Tehran, Tehran, Iran, in 2015, both in electrical engineering. He is currently working toward the Ph.D. degree with The University of Texas at Austin, Austin, TX, USA.
\end{IEEEbiographynophoto}
\vspace{-35pt}
\begin{IEEEbiographynophoto}{Zibo Chen} received
the B.S. degree from the Dalian University of Technology, Dalian, China, in 2016, and the M.Sc. degree from the Huazhong University of Science and Technology, Wuhan, China, in 2019, all in electrical engineering. He is currently working toward the Ph.D. degree with the University of Texas at Austin, Austin, TX, USA.
\end{IEEEbiographynophoto}
\vspace{-35pt}
\begin{IEEEbiographynophoto}{Chen Chen} received the B. S. degree from the Harbin Institute of Technology in electrical and automation engineering in 2020. He is currently working toward the Ph.D. degree with the University of Texas at Austin, Austin, Tx, USA.
\end{IEEEbiographynophoto}

\vspace{-35pt}
\begin{IEEEbiographynophoto}{Alex, Q. Huang} (Fellow, IEEE) was born in
Zunyi, China. He received the B.Sc. degree
from Zhejiang University, Hangzhou, China, in
1983, the M.Sc. degree from Chengdu Institute
of Radio Engineering, Chengdu, China, in 1986,
and the Ph.D. degree from Cambridge University, Cambridge, U.K., in 1992, all in electrical
engineering.
From 1992 to 1994, he was a Research Fellow with Magdalene College, Cambridge University. From 1994 to 2004, he was a Professor
with Bradley Department of Electrical and Computer Engineering, Virginia Polytechnic Institute and State University, Blacksburg, USA.
From 2004 to 2017, he was the Progress Energy Distinguished Professor of electrical and computer engineering with NC State University,
Raleigh, USA, where he established and led the NSF FREEDM Systems Engineering Research Center. Since 2017, he has been the
Dula D. Cockrell Centennial Chair of engineering with the University of
Texas at Austin, Austin, USA. Since 1983, he has been involved in the
development of modern power semiconductor devices and power integrated circuits. He fabricated the first IGBT power device in China, in
1985. He is the inventor of the emitter turn-off (ETO) thyristor, MCD,
IGPT, and ECT power devices. He developed the concept of Energy
Internet and the smart transformer-based energy router technology.
He has mentored and graduated more than 100 Ph.D. and master's
students, and has authored or co-authored more than 700 articles in
international conferences and journals. He has also been granted
more than 25 U.S. patents. His research interests include power electronics, power management microsystems, and power semiconductor
devices.
Dr. Huang is a Fellow of the National Academy of Inventors. He was
the recipient of the NSF CAREER Award, prestigious R\&D 100 Award,
MIT Technology Review's 2011 Technology of the Year Award, and 2019
IEEE Industry Applications Society Gerald Kliman Innovator Award and
2020 IEEE PELS R. David Middlebrook Achievement Award.
\end{IEEEbiographynophoto}
\vspace{100pt}
~
\end{document}